\documentclass{aa}   

\usepackage{graphicx}
\usepackage{txfonts}

\usepackage{gensymb}
\usepackage{wasysym}
\usepackage{caption}
\usepackage{subcaption}
\usepackage{threeparttable}
\usepackage{pdflscape}
\usepackage{rotating}
\usepackage{soul}

\usepackage{natbib}
\bibpunct{(}{)}{;}{a}{}{,}

\usepackage[dvipsnames]{xcolor}
\usepackage{hyperref}
\hypersetup{
    colorlinks=true,
    linkcolor=MidnightBlue,
    filecolor=blue,      
    urlcolor=cyan,
    citecolor=TealBlue
}

\begin{document}

   \title{Very-long-baseline interferometry study of the flaring blazar TXS\,1508+572 in the early Universe}

   \author{P. Benke\inst{1,}\inst{2}
        \and
          A. Gokus\inst{3}
        \and
          M. Lisakov \inst{4, 5, 1}
        \and
          L.~I.~Gurvits\inst{6, 7}
        \and
          F. Eppel\inst{2,}\inst{1}
        \and
          J. Heßd\"orfer\inst{2,}\inst{1}
        \and
          M. Kadler\inst{2}
        \and
          Y.~Y. Kovalev\inst{1}
        \and
          E. Ros\inst{1}
        \and
          F. R\"osch\inst{2,}\inst{1}
          }

   \institute{Max-Planck-Institut f\"ur Radioastronomie,
           Auf dem H\"ugel 69, D-53121 Bonn, Germany
         \and
            Julius-Maximilians-Universität Würzburg, Fakultät für Physik und Astronomie, Institut für Theoretische Physik und Astrophysik, Lehrstuhl für Astronomie, Emil-Fischer-Str. 31, D-97074 Würzburg, Germany
        \and
            Department of Physics \& McDonnell Center for the Space Sciences, Washington University in St. Louis, One Brookings Drive, St. Louis, MO 63130, USA
        \and
            Instituto de F\'{i}sica, Pontificia Universidad Cat\'{o}lica de Valpara\'{i}so, Casilla 4059, Valpara\'{i}so, Chile
        \and
            Astro Space Center of Lebedev Physical Institute, Profsouznaya 84/32, Moscow 117997, Russia   
        \and
            Joint Institute for VLBI ERIC, Oude Hoogeveensedijk 4, 7991 PD Dwingeloo, The Netherlands
        \and
            Faculty of Aerospace Engineering, Delft University of Technology, Kluyverweg 1, 2629 HS Delft, The Netherlands
        }

   \date{Received ; accepted }

 \abstract{High-redshift blazars provide valuable input to studies of the evolution of active galactic nuclei (AGN) jets and provide constraints on cosmological models. Detections at high energies ($0.1<\mathrm{E}<100$~GeV) of these distant sources are rare, but when they exhibit bright gamma-ray flares, we are able to study them. However, contemporaneous multi-wavelength observations of high-redshift objects ($z>4$) during their different periods of activity have not been carried out so far. An excellent opportunity for such a study arose when the blazar \object{TXS\,1508+572} ($z=4.31$) exhibited a $\gamma$-ray flare in 2022 February in the $0.1-300$~GeV range with a flux 25 times brighter than the one reported in the in the fourth catalog of the \textit{Fermi} Large Area Telescope.}
 {Our goal is to monitor the morphological changes, spectral index and opacity variations that could be associated with the preceding $\gamma$-ray flare in \object{TXS\,1508+572} to find the origin of the high-energy emission in this source. We also plan to compare the source characteristics in the radio band to the blazars in the local Universe ($z<0.1$). In addition, we aim to collect quasi-simultaneous data to our multi-wavelength observations of the object, making \object{TXS\,1508+572} the first blazar in the early Universe ($z>4$) with contemporaneous multi-frequency data available in its high state.}
 {In order to study the parsec-scale structure of the source, we performed three epochs of very-long-baseline interferometry (VLBI) follow-up observations with the Very Long Baseline Array (VLBA) supplemented with the Effelsberg $100$-m Telescope at $15$, $22$, and $43$\,GHz, which corresponds to $80$, $117$, and $228$\,GHz in the rest frame of \object{TXS\,1508+572}. In addition, one $86$\,GHz ($456$\,GHz) measurement was performed by the VLBA and the Green Bank Telescope during the first epoch.}
 {We present total intensity images from our multi-wavelength VLBI monitoring that reveal significant morphological changes in the parsec-scale structure of \object{TXS\,1508+572}. The jet proper motion values range from $0.12$~mas/yr to $0.27$~mas/yr, which corresponds to apparent superluminal motion $\beta_{\mathrm{app}}\approx14.3\,c$ -- $32.2\,c$. This is consistent with the high Lorentz factors inferred from the spectral energy distribution (SED) modeling for this source. The core shift measurement reveals no significant impact by the high-energy flare on the distance of the $43$-GHz radio core with respect to the central engine, that means this region is probably not affected by e.g., injection of new plasma as seen in other well-studied sources like CTA\,102. We determine the average distance from the $43$-GHz radio core to the central supermassive black hole to be $46.1\pm2.3~\mu$as, that corresponds to a projected distance of $0.32\pm0.02$~pc. We estimate the equipartition magnetic field strength $1$~pc from the central engine to be on the order of $1.8$~G, and the non-equipartition magnetic field strength at the same distance to be about $257$~G, the former of which values agrees well with the magnetic field strength measured in low to intermediate redshift AGN.}
 {Based on our VLBI analysis, we propose that the $\gamma$-ray activity observed in February 2022 is caused by a shock-shock interaction between the jet of \object{TXS\,1508+572} and new plasma flowing through this component. Similar phenomena have been observed, for example, in CTA\,102 in a shock-shock interaction between a stationary and newly emerging component. In this case, however, the core region was also affected by the flare as the core shift stays consistent throughout the observations.}


   \maketitle
%

\section{Introduction}
\label{intro}

Blazars, a subclass of radio-loud active galactic nuclei (AGN) whose jets point toward the observer \citep{1995PASP..107..803U}, are among the most luminous objects in the Universe. They emit radiation throughout the whole electromagnetic spectrum, and their spectral energy distribution (SED) shows a double-humped structure \citep{1995ApJ...444..567P}. In leptonic models, the low-energy hump, stretching from the radio to the ultraviolet, and sometimes even to the X-ray band, arises from synchrotron and the high-energy hump (X-rays to $\gamma$-rays) originates from inverse Compton scattering (IC). Depending on the origin of the seed photon field, the IC process can be either synchrotron self-Compton (SSC) or external Compton (EC) with seed photons from the broad-line region and/or the dusty torus. Alternatively, hadronic models (e.g., proton synchrotron) can also be responsible for the high-energy emission \citep[e.g.,][]{2002MNRAS.332..215A, 2013ApJ...768...54B}. High-redshift blazars that existed when the Universe was only $\sim1$~Gyr old already harbored supermassive black holes (SMBHs, $M_{\mathrm{BH}}\geq10^9 M_{\astrosun}$) in their central engines \citep{1995A&A...293L...1B, 2010MNRAS.405..387G}. However, it is not yet clear how these objects could have formed so early in the Universe, but studies by \citet{2008MNRAS.386..989J} and \citet{2013MNRAS.432.2818G} suggest that AGN feedback can boost accretion onto the central engine and accelerate black hole growth. Thus, investigating the properties of high-redshift blazars can help us to better understand SMBH and AGN evolution, as well as the intricacies of AGN feedback on their host galaxies \citep{2010A&ARv..18..279V}.

\begin{table*}[h!]
    \centering
        \caption{Summary of our VLBI observations.}
    \label{tab:obs}
    \begin{tabular}{c c c c}
    \hline\hline
       Epoch  & Array\tablefootmark{a} & Frequencies [GHz] & Comments \\
    \hline
        2022.24 & VLBA+Effelsberg & $15$, $22$, $43$ & Pt only recorded RCP \\ 
        2022.24 & VLBA+GBT & $86$ & Pt only recorded RCP\\
        2022.67 & VLBA+Effelsberg & $15$, $22$, $43$ & no Kp\\
        2023.05 & VLBA+Effelsberg & $15$, $22$, $43$ & no Hn, Kp \\
    \hline
    \end{tabular}
    \tablefoot{
    \tablefoottext{a}{Antenna names:  Br - Brewster, Eb - Effelsberg, Fd - Fort Davis, Hn - Hancock, Kp - Kitt Peak, La - Los Alamos, Mk - Mauna Kea, Nl - North Liberty, Ov - Ovens Valley, Pt - Pie Town, Sc - Saint Croix, GBT - Green Bank Telescope.}
    }
\end{table*}

\begin{figure*}[h!]
    \centering
    \includegraphics[width=\linewidth]{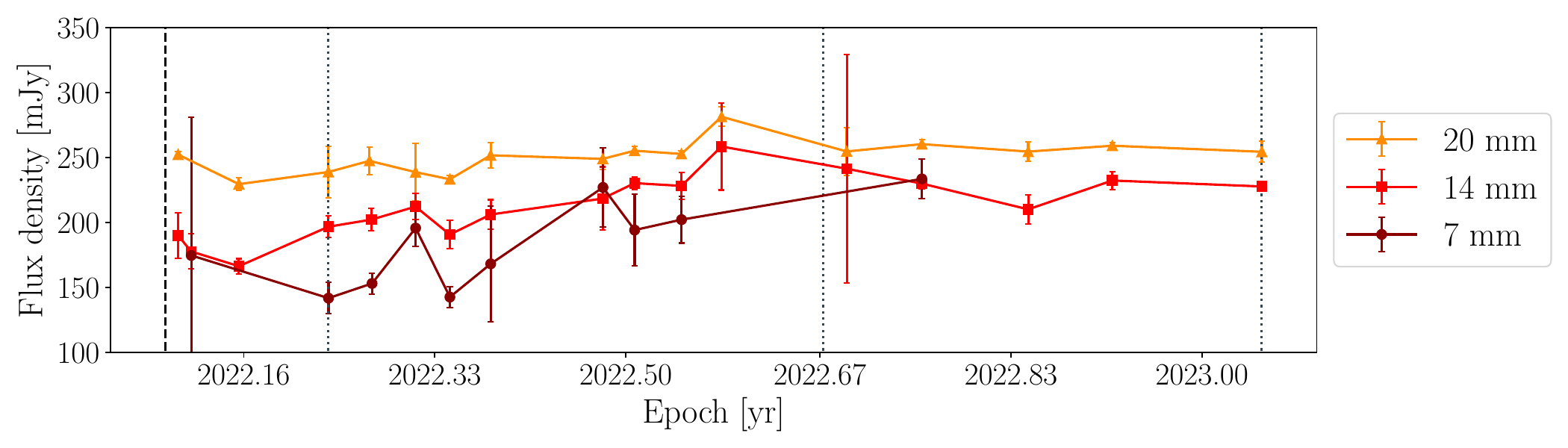}
    \caption{Radio light curve of \object{TXS\,1508+572} observed with the Effelsberg $100$-m telescope after the $\gamma$-ray flare (dashed black line). The VLBI observations are marked with black dotted lines.}
    \label{fig:ef_lc}
\end{figure*}

Blazars are the most abundant sources on the extragalactic $\gamma$-ray sky. While over half of the $5064$ sources contained in the fourth catalog of the \textit{Fermi} Large Area Telescope \citep[LAT,][]{2009ApJ...697.1071A} are blazars, only $33$ of these sources are categorized as high redshift objects \citep[$z>2.5$ cutoff set by the LAT,][]{2020ApJS..247...33A}. However, most of these high-redshift sources can only be detected by the LAT during flaring states \citep{2019ApJ...871..211P,2020ApJ...903..128K}, otherwise their steeply falling spectra prevent their detection at low states in the GeV energy range. This complicates obtaining quasi-simultaneous multi-wavelength observational data for studies of their broadband emission. The only example of such a quasi-simultaneous study to date is the case of the intermediate-redshift object \object{TXS\,0536+145} at $z = 2.69$ \citep{2014MNRAS.444.3040O}.

\object{TXS\,1508+572} (also called \object{GB6\,B1508+5714}, \object{J1510+5702}) is a high-redshift blazar at $z = 4.31$ \citep{1995MNRAS.273L..63H,2007AJ....134..102S}. On kiloparsec scales, the source shows a double-sided jet structure \citep{2022A&A...663A..44K} in the east-west direction. The first very-long-baseline interferometry (VLBI) image of \object{TXS\,1508+572} was published by \citet{1997A&A...325..511F}, and the resulting $5$\,GHz map shows an unresolved source structure with the synthesized beam of $\sim 5$\,mas. However, global VLBI observations at 5 and 8.4\,GHz revealed an optically thin jet component at 5~GHz and 8.4~GHz about 2\,mas south of the core \citep{VLBP-SOS-DCG-2011}. The AstroGeo Database\footnote{\url{http://astrogeo.org/vlbi_images/}} provides $8.7$\,GHz images of the source with a compact core-jet morphology and the jet oriented toward southwest. Kinematic analysis based on $4$ years of $8.6$\,GHz observations reveals a jet proper motion of $0.117\pm 0.078$~mas/yr \citep{2023AJ....165...69T}.

A strong $\gamma$-ray flare was detected in \object{TXS\,1508+572} on 2022 February 04 \citep{2022ATel15202....1G} with a $\gamma$-ray monitoring program following the high-$z$ blazar detection method described in \citet{2020ApJ...903..128K}. We have started a multi-frequency campaign across the electromagnetic spectrum to follow up this event with quasi-simultaneous observations \citep[see Paper I.,][]{paper1}. Since $\gamma$-ray observations lack the resolution required for determining the origin of the activity, we have initiated a multi-frequency VLBI monitoring to capture the evolution of the source morphology and, possibly, relate VLBI structural components to the observed high-energy activity. To the best of our knowledge, such an immediate follow-up observing campaign has never been carried before for any blazar at such a high redshift.

In Sect.~\ref{obs} we describe VLBI observations aimed to trace the source's morphological evolution during its high state. Sect.~\ref{analysis} describes the analysis of the observing data. Our results are discussed in Sect.~\ref{disc}, and we summarize our findings in Sect.~\ref{sum}. In this work we assume a $\Lambda$CDM cosmology with $H_0=70.7$~km~s$^{-1}$~Mpc$^{-1}$, $\Omega_{\Lambda}=0.73$, and $\Omega_{\mathrm{M}}=0.27$ \citep{2013PhRvD..88b3531S}. At the redshift of $z=4.31$, this corresponds to a scale of $6.9$~pc/mas, and a luminosity distance of $D_{L}\approx 40$\,Gpc.

\begin{figure*}[h!]
    \centering
    \begin{subfigure}{0.48\linewidth}
    \includegraphics[width=\linewidth]{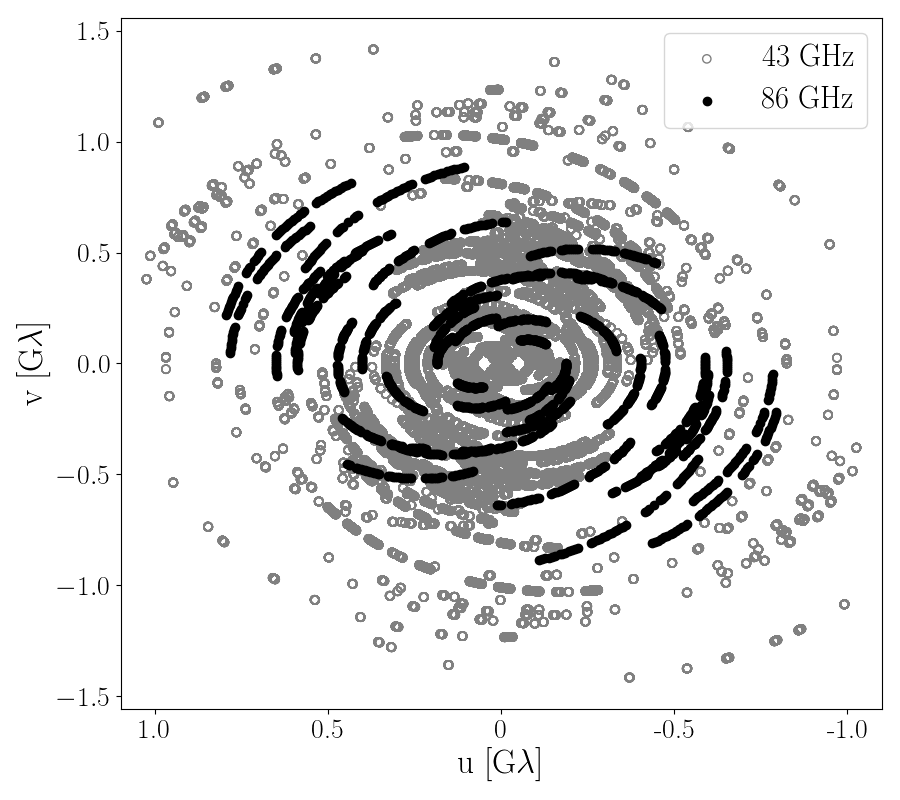}
    \end{subfigure}
    \begin{subfigure}{0.48\linewidth}
    \includegraphics[width=\linewidth]{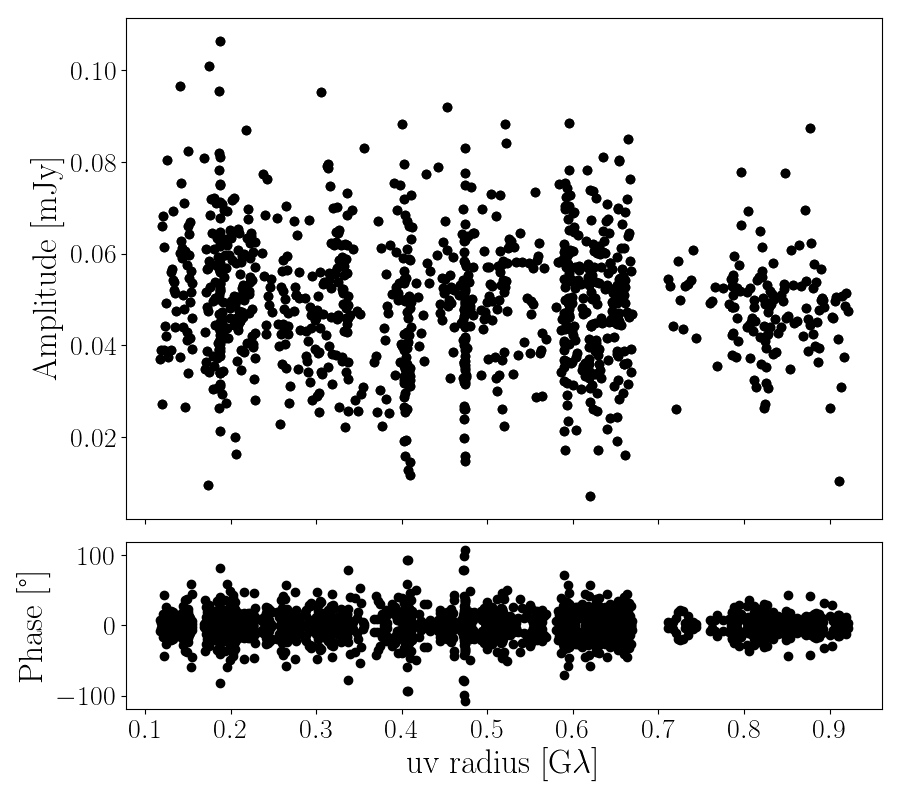}
    \end{subfigure}
    \caption{\textit{Left panel:} ($u$, $v$) coverage of the $43$-GHz VLBA and EF (grey) and the $86$-GHz VLBA and GBT (black) observations, plotting the fringe detections. \textit{Right panel:} Self-calibrated amplitude and phase of the $86$-GHz data set.}
    \label{fig:radplot}
\end{figure*}

\section{Observations and data reduction}
\label{obs}

To correlate the $\gamma$-ray activity with possible parsec-scale morphological changes in \object{TXS\,1508+572}, we requested three Very Long Baseline Array (VLBA) and Effelsberg observations at $15$, $22$, and $43$\,GHz, as well as an additional observation with the VLBA and the Green Bank Telescope (GBT) at $86$\,GHz (PIs: A. Gokus, M. Lisakov, project code: BG281).

Observations are summarized in Table~\ref{tab:obs}. The correlation was carried out using the VLBA DiFX correlator \citep{2007PASP..119..318D, 2011PASP..123..275D}, in $4$ subbands (intermediate frequencies or IFs) and two circular polarizations, each with $256$ spectral channels and a bandwidth of $128$~MHz. The integration time was $0.5$\,s for the $86$\,GHz observations and $1$\,s for all other observations. The source \object{1803+784} was used as a fringe finder calibrator in all experiments. 

We performed the calibration according to the standard recipes in the Astronomical Image Processing Software \citep[\texttt{AIPS},][]{2003ASSL..285..109G}. After loading the data to \texttt{AIPS} using \texttt{FITLD}, we applied parallactic angle and digital sampling corrections. Amplitude calibration was performed based on the system noise temperature and elevation dependent gain curves provided by the stations. Opacity corrections were also applied at this step based on the weather information recorded at each antenna site. We used \texttt{FRING} to determine phase delay and rate solutions, and applied them both to the calibrator and target using \texttt{CLCAL}. It is worthwhile to highlight the importance of the GBT in the $86$-GHz observation, because sufficiently high signal-to-noise ratio (S/N) fringe detections ($\mathrm{S/N}>5$) were only found on GBT baselines. These solutions were applied to the data and the fringe fit was repeated setting delay and rate windows of $200$~nsec and $200$~mHz, and lowering the S/N cutoff to $3.7$.

We corrected the flux density scale based on the quasi-simultaneous single-dish observations at $20$, $14$, and $7$~mm of \object{TXS\,1508+572} with the Effelsberg $100$-m telescope\footnote{The data is available at: \url{https://telamon.astro.uni-wuerzburg.de/}.} (see Fig.~\ref{fig:ef_lc}). Observations and data reduction was carried out as described in \citet{2024arXiv240106296E}. The data was then averaged in frequency with \texttt{SPLIT} and written out for imaging. 

\begin{figure*}[h!]
    \centering
    \includegraphics[width=\linewidth]{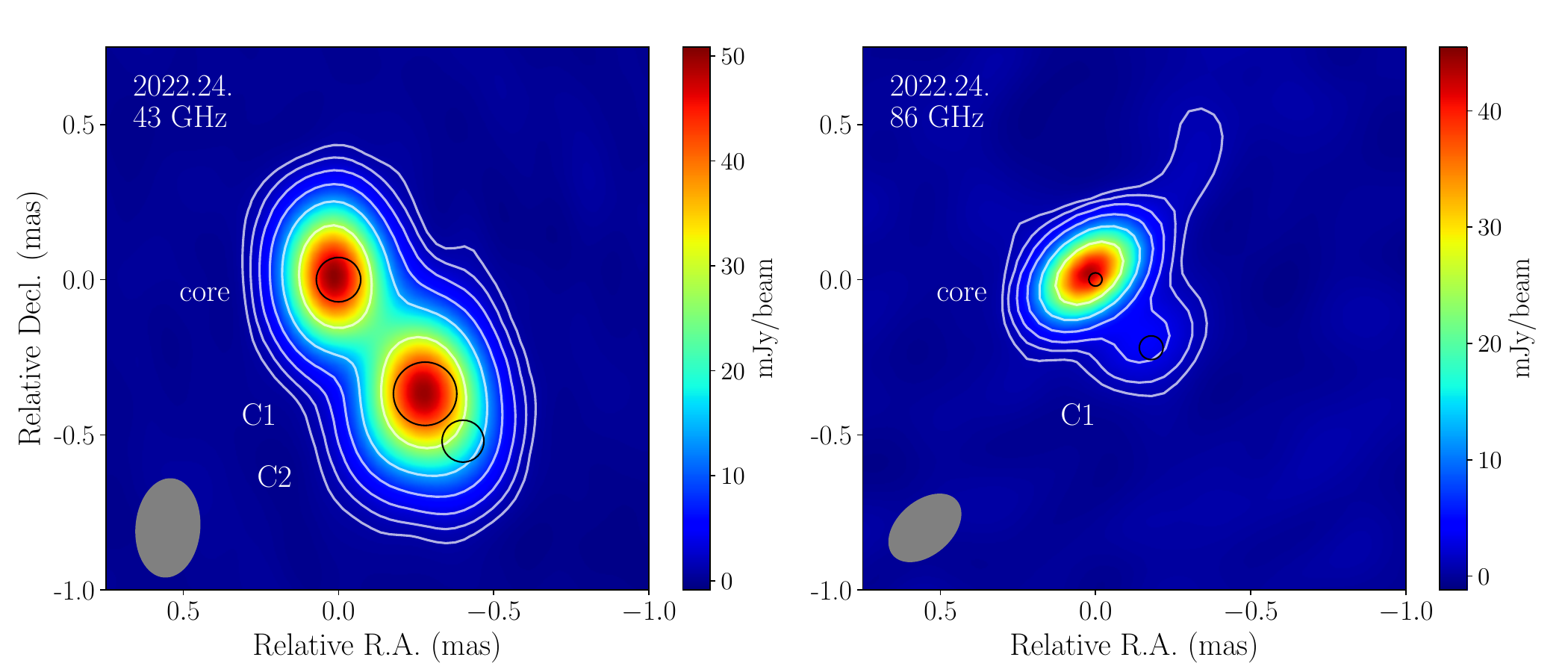}
    \caption{$43$ and $86$\,GHz image of \object{TXS\,1508+572} observed with the VLBA+EF ($43$\,GHz, left) and VLBA+GBT ($86$\,GHz, right). Lowest contours are at 0.53 and 1.37~mJy/beam, and increase as a factor of two. Positions of the \texttt{modelfit} components are overlaid as black circles.}
    \label{fig:86}
\end{figure*}

Hybrid imaging was performed in \texttt{Difmap} \citep{1997ASPC..125...77S} with iterating \texttt{clean} and phase and amplitude self-calibration with decreasing solution interval ranging from $180$ to $1$~minutes. Due to the low quality of the $86$-GHz data (see Fig.~\ref{fig:radplot}), the imaging was carried out in two ways: assuming the same source structure as seen at lower frequencies, that is loading the $43$-GHz clean windows to image the $86$-GHz data; and assuming a core-dominated source. The latter yields a better map, because we do not detect any significant emission at the location of the $43$-GHz jet component. Resulting maps are displayed in Fig.~\ref{fig:cln} and ~\ref{fig:86} and their properties are tabulated in Tab.~\ref{tab:cln}.

\section{Data analysis}
\label{analysis}

\subsection{Model fitting of individual source components}

 To be able to track individual components, as well as their brightness and kinematic evolution, we modeled the source structure with delta and elliptical Gaussian components using the \texttt{modelfit} command in \texttt{Difmap} (see Tab.~\ref{tab:comps}). The source is usually well modeled with four components representing the core, two jet components, and the lobe. To measure the jet components' proper motion, $\mu$, we calculated the angular separation between the core and jet components and fitted the obtained values using linear regression. Results from our kinematic measurements are shown in Fig.~\ref{fig:kin}. The obtained values of proper motion are $\mu_{15, C1} = 0.19 \pm 0.12$\,mas/yr, $\mu_{15, C2} = 0.27\pm0.16$\,mas/yr, $\mu_{22, C1} = 0.20 \pm 0.09$\,mas/yr, $\mu_{22, C2} = 0.21\pm0.10$\,mas/yr, $\mu_{43, C1} = 0.12\pm0.20$, and $\mu_{43, C2} = 0.15 \pm 0.29$\,mas/yr at $15$, $22$, and $43$\,GHz, respectively. These values are somewhat higher than the value $\mu_{8} = 0.117 \pm 0.078$~mas/yr from \cite{2023AJ....165...69T} based on observations at $8$\,GHz (see Sect.~\ref{disc2}). In Sect.~\ref{mf_kin} we discuss the multi-frequency kinematic analysis of the two jet components.
 
 We calculated the brightness temperature in the source frame, $T_{\mathrm{b, obs}}$, the following way:
 \begin{equation}
    T_{\mathrm{b, obs}} [\mathrm{K}] = 1.22\times 10^{12} \Bigg(\frac{S_{\nu}}{\mathrm{Jy}}\Bigg) \Bigg(\frac{\nu}{\mathrm{GHz}}\Bigg)^{-2} \Bigg(\frac{b_{\mathrm{comp}}}{\mathrm{mas}}\Bigg)^{-2} (1+z),
\end{equation}
where $S_{\nu}$ is the flux density of the components, $\nu$ is the observing frequency and $b_{\mathrm{comp}}$ is the size (full width at half maximum, FWHM) of the component. If a component is not resolved according to the resolution limit calculated based on Eq. $2.$ from \citet{2005AJ....130.2473K}, we give an upper limit on the component size, and calculate a lower limit for $T_{\mathrm{b, obs}}$. These values are listed in Tab.~\ref{tab:comps}. Taking Doppler boosting into account with a Doppler factor of $\delta\approx20$, all values are below the equipartition brightness temperature of $T_{\mathrm{eq}}\approx5\times10^{10}$~K \citep{1994ApJ...426...51R}.

\subsection{Spectral index maps and core shift measurements}
\label{spec_cs}

To create spectral index maps, we re-mapped the image pairs in order for them to have a similar ($u$, $v$) range, pixel size, and restoring beam size. Images were then aligned on the optically thin jet components using 2D cross-correlation. Spectral index maps are displayed in Fig.~\ref{fig:spix}, and spectra of the core and jet components are shown in Fig.~\ref{fig:tb}. In the Blandford--Königl jet model \citep{1979ApJ...232...34B}, the VLBI core represents the $\tau=1$ optical depth to synchrotron radiation, whose geometry is frequency dependent. As a result of this, higher frequency observations probe the regions closer to the central engine, and one can extrapolate the location of the SMBH based on the relative frequency-dependent shifts of the optically thick core determined from pairs of images at different frequencies  \citep{1984ApJ...276...56M, 1998A&A...330...79L}. For our analysis, we use the $43$\,GHz core as a reference. Our core shift measurement is carried out the same way as described in \citet{2012A&A...545A.113P}, aligning the images using the shifts obtained from 2D cross-correlation, and measuring the difference between the core positions of consecutive frequency pairs. The alignment error was assumed to be half of the pixel size of a given frequency pair, and the core position errors were estimated based on the $\chi^2$ minimization method described in \citet{1976ApJ...208..177L}. Under the assumption of equipartition, conserved magnetic flux, as well as that both the particle density and the magnetic field decrease with the distance from the central engine \citep{1998A&A...330...79L}, we can then measure the apparent distance between the VLBI core and the jet apex as:

\begin{figure*}[h!]
    \centering
    \includegraphics[width=0.9\linewidth]{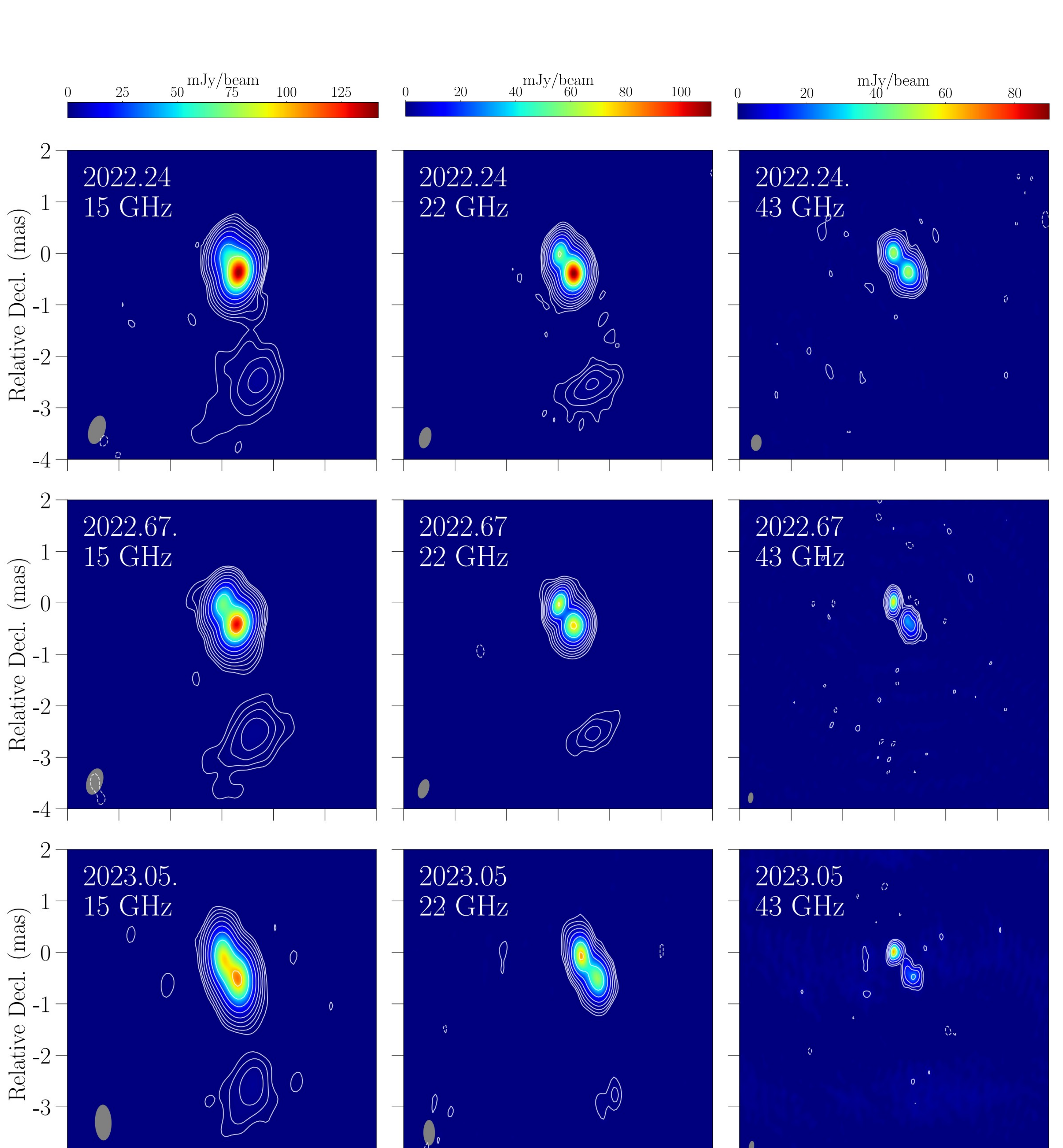}
    \caption{Clean images of \object{TXS\,1508+572} at $15$, $22$, and $43$\,GHz. The observations were carried out in $2022$ March, September, and in $2023$ January (see Table~\ref{tab:obs}). Image properties are summarized in Table~\ref{tab:cln}. Contours and colors represent the brightness distribution of the parsec-scale structure of the object. The images are aligned based on the core shift measurement.}
    \label{fig:cln}
\end{figure*}

\noindent
\begin{equation}
\Delta r_{\mathrm{core}} [\mu as] = r_0 \Bigg[\Bigg(\frac{\nu}{43 \mathrm{GHz}}\Bigg)^{-1/k_{\mathrm{r}}} -1 \Bigg],
\end{equation}
where $r_{0}$ is the distance between the jet apex and the $43$\,GHz core. $k_{\mathrm{r}}=1$ corresponds to a conical jet width profile. The results of our core shift analysis are shown in Fig.~\ref{fig:cs}.

Based on the core shift measurement, we estimate the magnetic field strength using the methods described in \citet{1998A&A...330...79L}, \citet{2005ApJ...619...73H}, \citet{2015MNRAS.451..927Z} and \citet{2017MNRAS.468.4478L}. Magnetic field strength measurements in high-redshift AGN can provide important insights into SMBH accretion and jet launching, such as in the case of the $z=3.40$ blazar, OH\,471 \citep{2024A&A...685L..11G}.
First, we calculate $\Omega_{\mathrm{r, \nu}}$, the shift in parsec per unit $1/\nu$ difference:
\begin{equation}
    \Omega_{\mathrm{r \nu}} [\mathrm{pc\,GHz^{k_{\mathrm{r}}}}] = 4.85\times10^{-9} \frac{\Delta r_{\mathrm{core}}D_{\mathrm{L}}}{(1+z)^2}  \frac{\nu_1^{1/k_{\mathrm{r}}} \nu_2^{1/k_{\mathrm{r}}}}{\nu_2^{1/k_{\mathrm{r}}}-\nu_1^{1/k_{\mathrm{r}}}} ,
\end{equation}
where $\Delta r_{\mathrm{core}}$ is the core shift between frequencies $\nu_1$ and $\nu_2$. To derive an upper limit on the magnetic field strength, we calculate the core shift and $\Omega_{\mathrm{r \nu}}$ between $15$ and $43$\,GHz.

The magnetic field strength $1$~pc from the jet apex under the equipartition assumption is calculated as \citep{2015MNRAS.451..927Z}:
\begin{equation}
    B^{\mathrm{eq}}_{\mathrm{1pc}} [G] \approx 0.025 \Bigg[\frac{\Omega^{3k_{\mathrm{r}}}_{r\nu} (1+z)^{3}}{\delta^{2} \phi \sin^{3k_{\mathrm{r}}-1}\theta} \Bigg]^{\frac{1}{4}},
\end{equation}
where $\phi$ is the intrinsic opening angle and $\theta$ is the viewing angle. For the viewing angle we adopted the same value as the one used for the SED fit in Paper I, $\theta=1/20$~rad. We calculated the apparent half opening angle based on the size of the \texttt{modelfit} components as $\phi_{\mathrm{app}}=\arctan[{(b_{\mathrm{jet}}-b_{\mathrm{core}})/d}$], which are the sizes of the jet, $b_{\mathrm{jet}}$, and core components, $b_{\mathrm{core}}$, and the distance between these components, $d$. We did this for all observations and adopted $\phi_{\mathrm{app}}$ as the average of these values, $(9\pm2)$\degree. The intrinsic full opening angle was calculated as $\phi=2\arctan(\tan{\phi_{\mathrm{app}}}\sin{\theta})$, which is $(0.9\pm0.2)$\degree.

We also calculate the magnetic field strength $1$~pc from the central SMBH, without assuming equipartition \citep{2015MNRAS.451..927Z}:
\begin{equation}
    B^{\mathrm{non-eq}}_{\mathrm{1pc}} [\mathrm{G}] \approx \frac{3.35\times10^{-11}  D_L \Delta r_{\mathrm{core}} \delta \tan{\phi}}{(\nu_1^{-1} - \nu_2^{-1})^5 [(1+z) \sin{\theta}]^3 F_{\nu}^2},
\end{equation}  
where $F_{\nu}$ is the flux density in the flat part of the spectrum.

The equipartition magnetic field strengths, $B^{\mathrm{eq}}_{\mathrm{1pc}}$, derived for the three considered epochs are the following: $1.91\pm0.11$~G, $1.57\pm0.17$~G, and $1.79\pm0.16$~G. The $B^{\mathrm{non-eq}}_{\mathrm{1pc}}$ values for the same epochs are $324.7\pm156.3$~G, $104.9\pm82.7$~G, and $341.9\pm225.2$~G, respectively. 

In order to compare $B^{\mathrm{eq}}$ to that of M\,87 on horizon scales \citep{2021ApJ...910L..13E}, we extrapolate the average $B^{\mathrm{eq, aver}}=0.79\pm0.04$~G to $5~r_{\mathrm{g}}$ or $0.0036$~pc projected distance. The gravitational radius is calculated as $r_{\mathrm{g}} = GM_{\mathrm{BH}}/c^2$, where $G$ is the gravitational constant and $M_{\mathrm{BH}}$ is the black hole mass assumed to be $1.5\times10^{10}~\mathrm{M}_{\odot}$ based on the SED fit parameters in \citet{paper1}. We measure $B^{\mathrm{eq}}_{\mathrm{5r_g}}=487.4\pm33.0$~G, that is significantly larger than the $B\sim1-30$~G reported for M\,87 \citep{2021ApJ...910L..13E}.

\section{Discussion}
\label{disc}

\subsection{Change in source morphology related to the $\gamma$-ray flare}
\label{disc1}

Our VLBI observing campaign on \object{TXS\,1508+572} started after the detection of a bright $\gamma$-ray flare on 2022~February~04. Our observations at $15$, $22$, and $43$\,GHz correspond to $80$, $117$, and $228$\,GHz in the rest frame of the source given its redshift $z=4.3$. Our observation at $86$\,GHz ($456$\,GHz in the rest frame of the object) carried out with the VLBA and the GBT reaches frequency ranges similar to what is currently available with the Event Horizon Telescope \citep{eht19a}. At these observing frequencies we expect to probe the jet components in the optically thin regime.

Hybrid images from our multi-frequency observations are shown in Fig.~\ref{fig:cln}. We identify the core to be the northeastern component, as it has a flat spectrum (see Fig.~\ref{fig:spix}), and it is more compact than the southwestern jet component (Tab.~\ref{tab:comps}). We denote the southernmost faint, diffuse component as the lobe. Its position coincides with the jet detected at $8.6$\,GHz by \citet{2023AJ....165...69T}, and it is most likely the remnant of previous activity of the blazar. $1.7$ and $4.8$-GHz very high resolution \textit{RadioAstron} space VLBI observations reveal a similar source structure to what we observe in our images (L.~I. Gurvits, priv. comm.). The $43$ and $86$\,GHz images of \object{TXS\,1508+572} show only a compact core-jet morphology. Comparing the structure seen at $144$~MHz with LOFAR \citep{2022A&A...663A..44K} and at $1.4$\,GHz with the VLA \citep{2004ApJ...600L..23C} to our high-frequency images, we note a difference in the jet orientation on kiloparsec and parsec scales. This projection effect, when the intrinsic bending of the jet is amplified via beaming, is commonly seen in AGN \citep{1988ApJ...328..114P, 1993ApJ...411...89C}.

\subsection{Kinematics at high redshifts}
\label{disc2}

While VLBI observations can measure intrinsic characteristics and kinematic properties of AGN jets, investigating high-redshift sources is challenging due to several reasons. As a result of their Doppler-boosted emission, blazars with compact core-jet structures tend to dominate flux limited AGN samples at any given redshift. In the case of AGN oriented at large angles to our line of sight, however, the steep-spectrum jet emission is too weak to be detected at high rest frame frequencies \citep{2000pras.conf..183G}. In addition, due to the time dilation caused by the expansion of the Universe, component movements are visible only on longer time intervals. The first jet proper motion measurements at $z>5$ were performed by \citet{2015MNRAS.446.2921F} with the cadence of $7$~yrs or $1.17$~yrs in the rest frame of the target, \object{J1026+2542}. While high-redshift AGN are being targeted by VLBI observations more frequently \citep[][and references therein]{2022ApJS..260...49K}, kinematic analysis is only available for a small subset of them \citep[][and references therein]{1997A&A...325..511F, 2010A&A...521A...6V, 2015MNRAS.446.2921F, 2018MNRAS.477.1065P, 2020NatCo..11..143A, 2020SciBu..65..525Z, 2022ApJ...937...19Z, 2023PASA...40....4G, 2023IAUS..375...86G}. Expanding this sample is crucial to widen our knowledge on the evolution of black hole jets. Kinematic analysis enable us to measure the bulk Lorentz factor and Doppler boosting factors, as well as the jet viewing angle. In addition, these observational data can be used to constrain SED model parameters (see Paper I), and can also be compared with the characteristics of local AGN. While high-redshift sources show only mildly relativistic apparent speeds \citep{2022MNRAS.511.4572A}, AGN in the local Universe exhibit a much wider range of jet speeds, and often show superluminal motion \citep{2021ApJ...923...30L}.

\begin{figure}[h!]
    \centering
    \includegraphics[width=\linewidth]{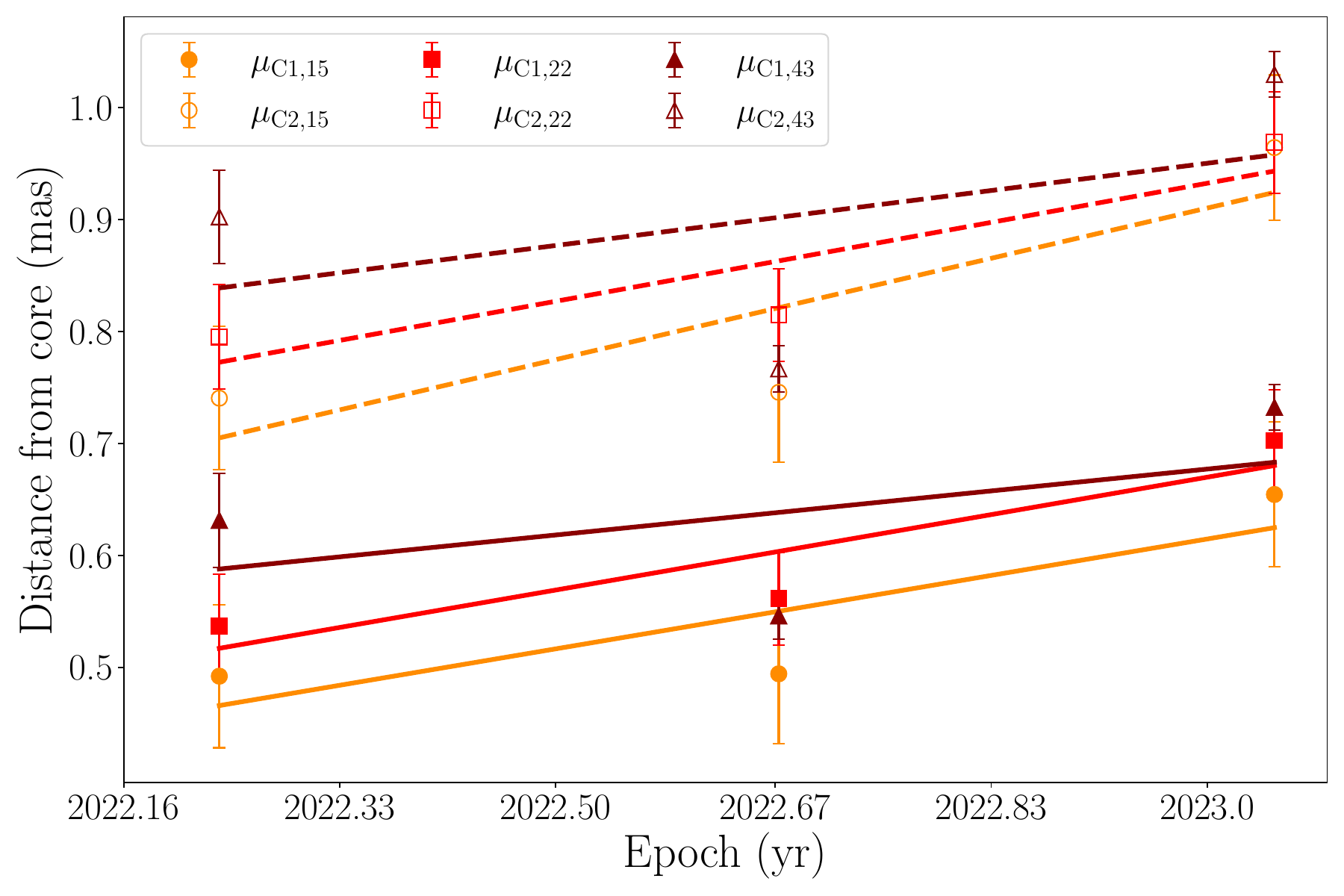}
    \caption{Kinematics of the jet component at $15$ (orange circles), $22$ (red squares), and $43$\,GHz (dark red triangles) following the $\gamma$-ray flare. Jet speeds are listed in the text in Sect.~\ref{analysis}. The positional errors are assumed to be one fifth of the beam minor axis. Note the systematic offset of the position of both jet components at the higher frequencies, which is roughly consistent with the core shift determined in Section~\ref{disc3}.}
    \label{fig:kin}
\end{figure}

The follow-up of the flaring activity in \object{TXS\,1508+572} spans $0.82$~yrs ($0.15$~yrs at $z = 4.31$), and reveals morphological changes on monthly timescales. This is most evident at $43$\,GHz, where the compact jet (southwest component in Fig.~\ref{fig:cln}) observed at $2022.24$ becomes fainter and more diffuse with time. Such changes on short timescales after a high-energy flare were not observed in high-redshift sources thus far. The results from our kinematic analysis, based on the \texttt{modelfit} parameters of the jet components in the three epochs of observations are shown in Fig.~\ref{fig:kin}. The measured jet speeds ranging between $\mu=0.12$ and $\mu=0.27$\,mas/yr correspond to apparent superluminal speeds of $\beta_{\mathrm{app}}\approx14.3-32.2$~$c$, where $\beta_{\mathrm{app}}=\mu D_{\mathrm{L}}/(1+z)$ is measured in the units of $c$. These jet speeds are higher than the value of $\mu_{8} = 0.117 \pm 0.078$~mas/yr presented in \citet{2023AJ....165...69T} based on $8$-GHz observations between $2017$ and $2021$. The discrepancy between the two kinematic measurements can be explained not only by the different time range and frequency coverage, but essentially by the fact that our and the cited kinematic estimates by \cite{2023AJ....165...69T} involve different structural components. Due to opacity effects (see Sect.~\ref{disc3}) and a higher angular resolution achieved in our higher frequency observations, we detect innermost components not distinguishable at $8$~GHz. The position of the jet component identified in \citet{2023AJ....165...69T} corresponds to the lobe component in our observations. It is therefore not surprising that the diffuse southern component at about $2$~mas from the core demonstrated a lower apparent velocity in the study by \cite{2023AJ....165...69T} even if the jet did not change its orientation along the stream relative to the line of sight. However, the inner jet and the diffuse lobe might also be oriented at different angles to the line of sight, resulting in a different apparent speed.

 Our kinematic measurement is consistent with the high Lorentz factor of $\Gamma=11$ obtained from modeling the SED during a quiescent state of \object{TXS\,1508+572} \citep{2017ApJ...837L...5A}, as well as a Lorentz factor of $20$ used to model the source during the flaring state in our Paper I. Brightness temperatures on the order of $10^9-10^{12}$~K (see Tab.~\ref{tab:comps} and the lower panel of Fig.~\ref{fig:tb}) also suggest Doppler-boosted emission. These values measured at high rest frame frequencies are comparable to the ones measured by \textit{RadioAstron}, $(5.15\pm2.1)\times10^{10}$~K at $1.67$\,GHz and $(2.15\pm3.3)\times10^{12}$ at $4.84$\,GHz, at extremely high angular resolution (L.~I. Gurvits, priv. comm.). In addition, apparent motion values agree well with expectations for high-redshift AGN based on the apparent proper-motion–redshift ($\mu-z$) relation 

\begin{figure}[h!]
    \centering
    \includegraphics[width=\linewidth]{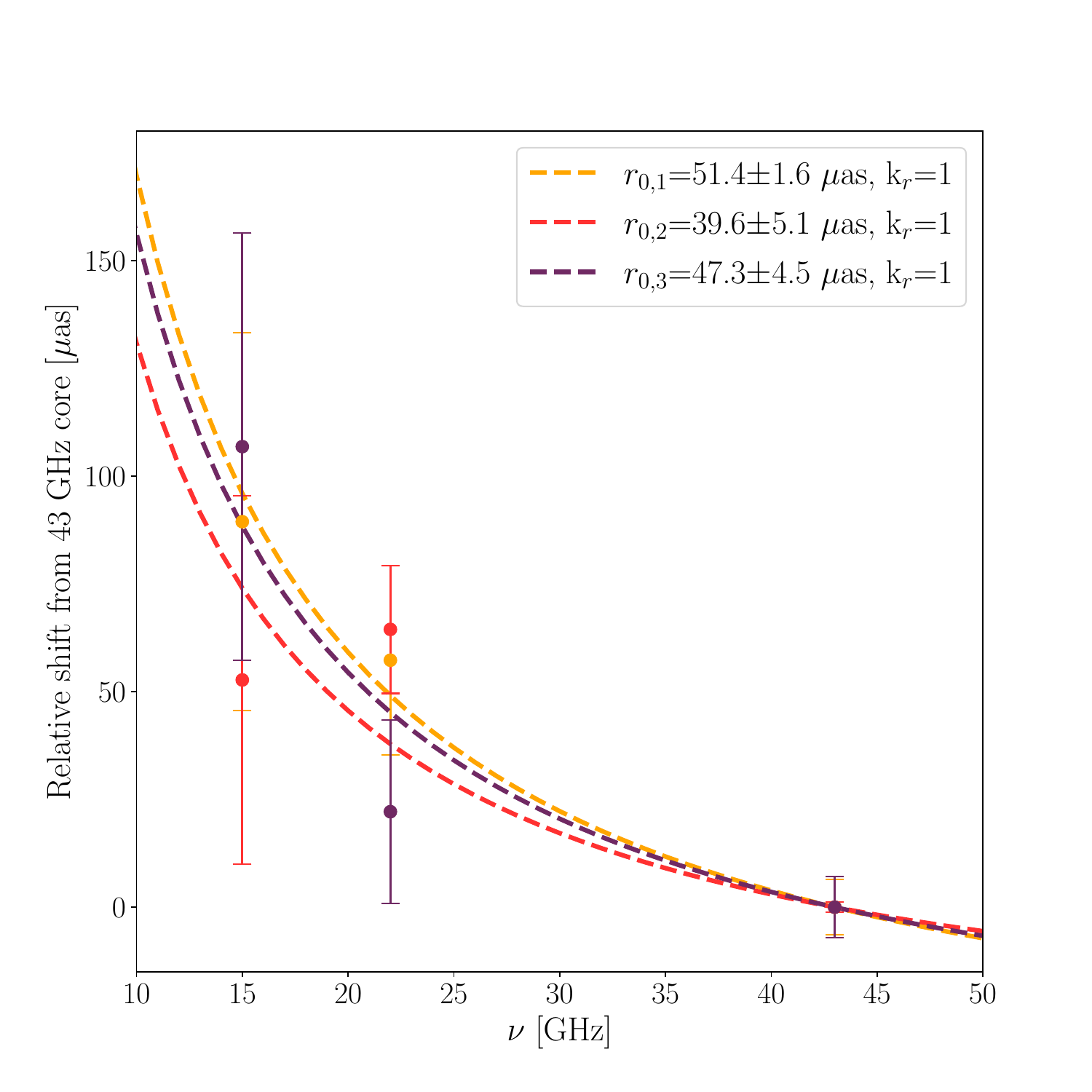}
    \caption{Evolution of the core shift measured throughout the first (gold), second (red), and third (purple) VLBI epochs. The distance to the jet apex ($r_{0}$) is shown in the legend for each of the three epochs.} 
    \label{fig:cs}
\end{figure}
 
 \noindent
\citep{1988ApJ...329....1C, 1994ApJ...430..467V, 1999NewAR..43..757K, 2015MNRAS.446.2921F, 2022ApJ...937...19Z}.  Comparing our measurements to the ($\mu-z$) relation based on $\leq15$\,GHz VLBI data in Fig.~$2$ of \citet{2022ApJ...937...19Z}, we find that all our jet proper motion values are consistent within the error bars with maximum apparent velocities assuming $\Gamma=20$. However, proper motion measurements in the $z = 4.33$ \citep{2001AJ....121.1799P} quasar \object{J2134$-$0419} reveal a significantly slower jet speed of $\mu = 0.035\pm0.023$\,mas/yr at $5$\,GHz. The difference, again, might be explained with the different observing frequencies, or with the high state during which \object{TXS\,1508+572} was observed, while \object{J2134$-$0419} shows no clear signs of flux density variability in the period studied.

Based on a simple linear fit to the distance from the radio core to the C1 and C2 components, we suggest that the ejection of the jet components fell around $\sim2016-2019$. Indeed, the source showed an elevated state in the GeV range and was detected with about $3\sigma$ significance during $2018-2020$. As a result of this, we suggest that the jet was not ejected during the current flaring activity. This, together with the cross-identification of the jet component in \citet{2023AJ....165...69T} to the lobe component in our images, we suggest that the $8$-GHz jet was ejected at an earlier epoch than the inner jet detected in our observations, which could explain the discrepancy in the proper motion values.

We have identified both components C1 and C2 at all frequencies and we also have measured the relative position of the apparent cores at different frequencies. Using these measurements together we have produced a combined kinematic plot presented in Fig.~\ref{fig:mfkin}. These combined data support a non-linear relation. However, with only three epochs it is impossible to distinguish between such options as accelerated motion \citep{2015ApJ...798..134H, 2022ApJS..260...12W} and moving of the apparent jet base \citep{2015ApJ...807L..14N, 2017MNRAS.468.4478L, 2019MNRAS.485.1822P}.

One argument for the latter option comes from the coordinated motion of components C1 and C2, that is they both appear closer to the apparent core at the second epoch. At the same time, these two components are casually disconnected, since they are located tens of parsecs away from each other and unlikely experience the same variation of their apparent velocity simultaneously. It brings us to a conclusion, that the reference point, that is the apparent core, might be moving itself. Such movement, indeed, is expected for the apparent core if denser plasma is flowing through it.

In this case, the second epoch might showcase the apparent core to be located more downstream, which made the distance to components C1 and C2 shorter. This can be explained by an increase of plasma density in the jet in 2022.67, possibly associated with the preceding $\gamma$-ray flare.

\subsection{Core shift evolution}
\label{disc3}

\begin{figure}[h!]
    \centering
    \includegraphics[width=\linewidth]{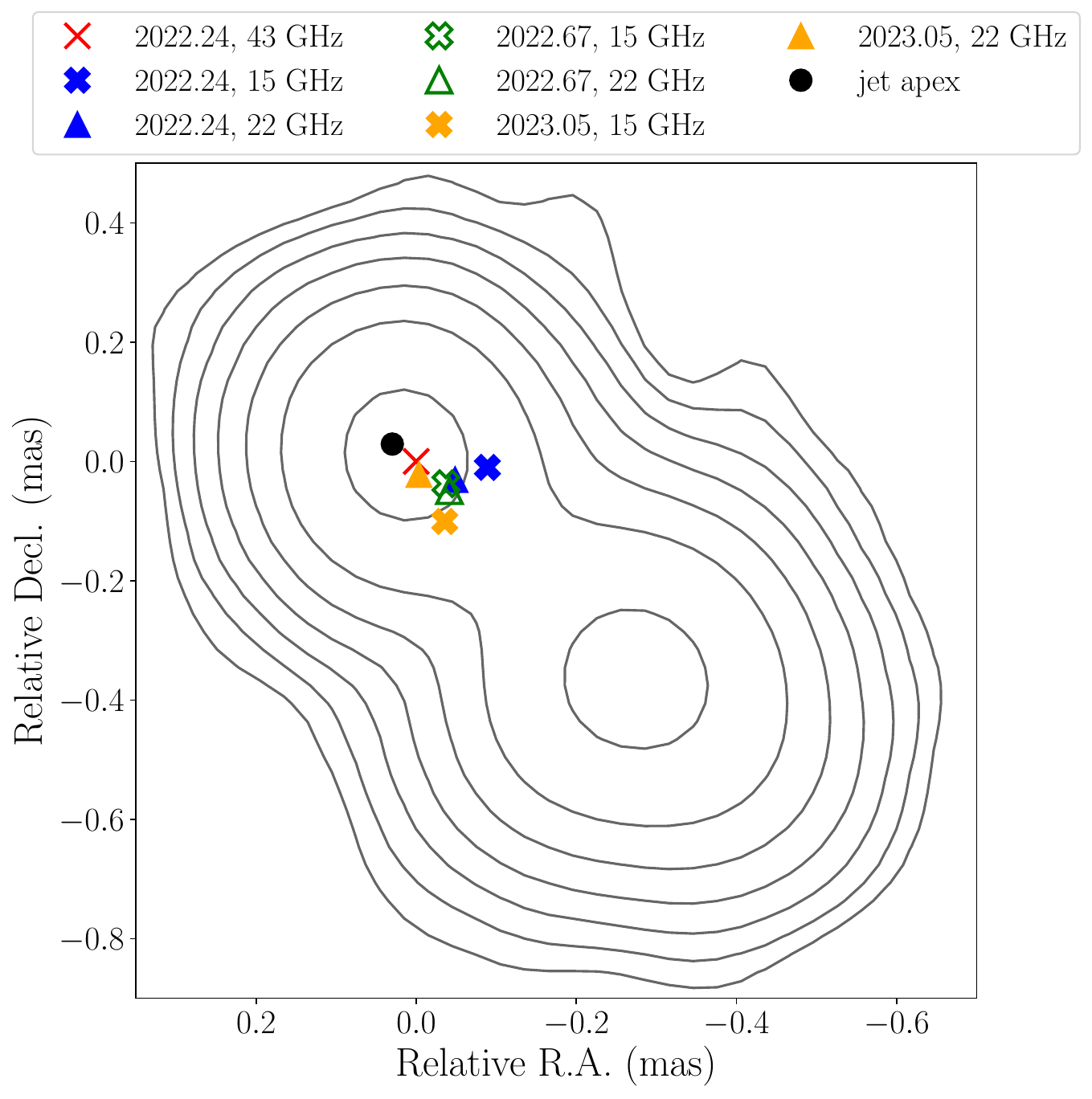}
    \caption{Location of the radio core with respect to the $43$-GHz one, overlaid on the $43$-GHz map from the first observing epoch. Error bars are omitted for clarity. The average location to the central supermassive black hole is marked with a black dot.}
    \label{fig:cs_map}
\end{figure}

The core shift measurements alone (see Fig.~\ref{fig:cs}), described in Sect.~\ref{spec_cs}, reveal no significant evolution in the distance to the jet apex subsequent to the high-energy flare. Our fits are consistent with a conical jet profile. The average distance of the $43$-GHz core to the central engine is $46.1\pm2.3~\mu$as corresponding to a projected distance of $0.32\pm0.02$~pc (see Fig.~\ref{fig:cs_map}). However, if we consider the core shift measurements together with the kinematics presented in Sect.~\ref{disc2}, we see a coherent picture regarding the second epoch. Displacement of the apparent core not only affects single-frequency kinematics, but can also affect single-epoch core shift measurements. Indeed, since apparent cores at different frequencies are separated by several parsecs along the jet, a moving feature displaces them non-simultaneously and possibly by different amounts \citep{2019MNRAS.485.1822P}. This behavior explains both kinematics and core shift measurements in a coherent manner. Unfortunately, large errors do not let us investigate this quantitatively.

Based on the core shift measurement, we derived the equipartition magnetic field strengths $1$~pc from the SMBH \citep{2015MNRAS.451..927Z}, which are $1.91\pm0.11$~G for the first, $1.57\pm0.17$~G for the second, and $1.79\pm0.16$~G for the third epoch. These magnetic field strengths are close to the ones derived for the MOJAVE sample \citep{2012A&A...545A.113P} and by \citet{2014Natur.510..126Z} for AGN below redshift $2.43$ and $2.37$, respectively. Non-equipartition magnetic field strengths $1$~pc from the SMBH are $324.7\pm156.3$~G, $104.9\pm82.7$~G, and $341.9\pm225.2$~G for the first, second, and third epochs, respectively. Extrapolating the average $B^{\mathrm{eq}}_{\mathrm{1pc}}$ to $5 r_{\mathrm{g}}$, we measure $B^{\mathrm{eq}}_{\mathrm{5r_g}}=487.4\pm33.0$~G, a value significantly larger than the $B\sim1-30$~G reported for M\,87 on horizon scales \citep{2021ApJ...910L..13E}.

\begin{figure}[h!]
    \centering
    \includegraphics[width=\linewidth]{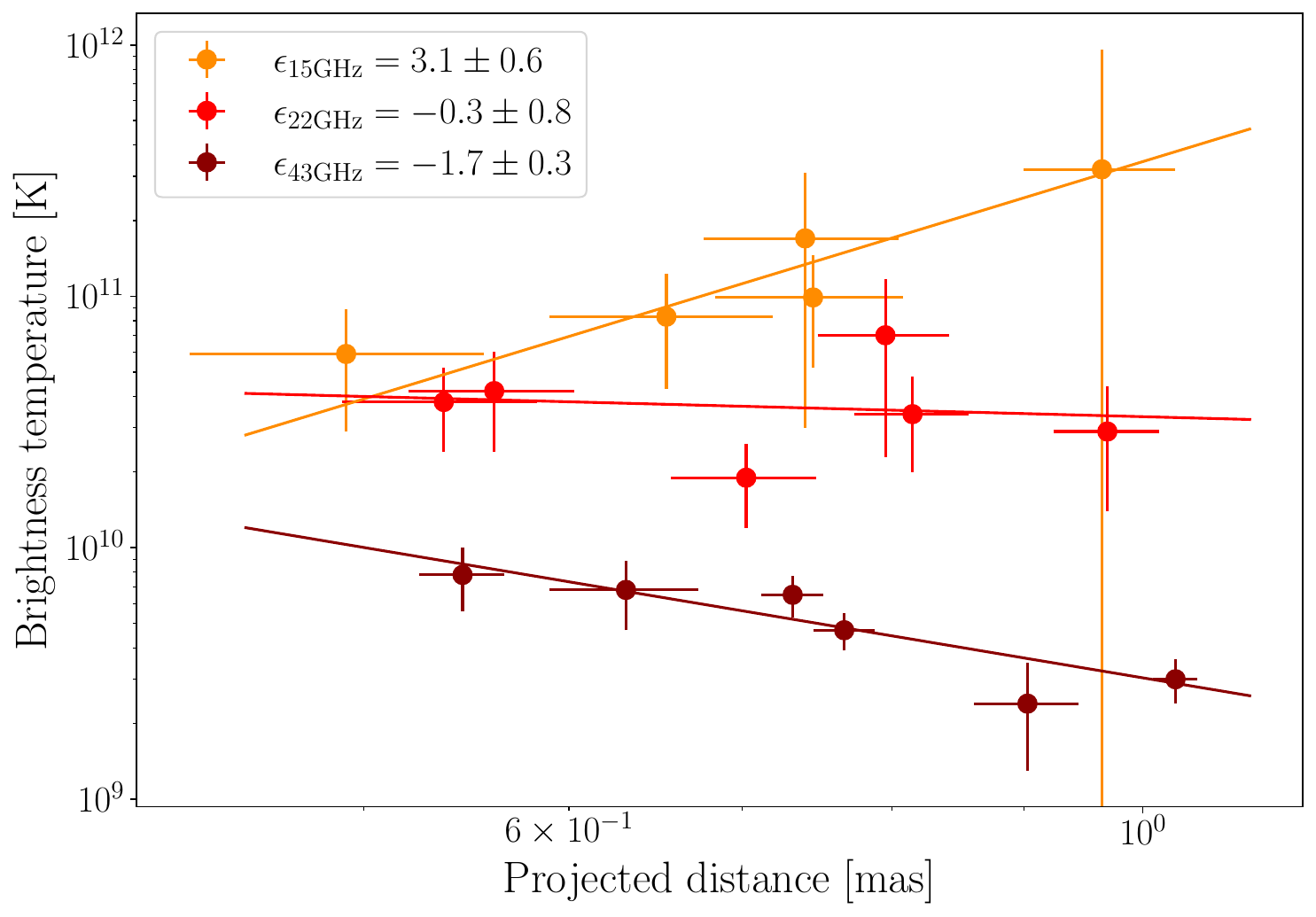}
    \caption{Jet brightness temperature as a function of projected distance from the core. Solid lines represent power-law fits to the $T_{\mathrm{b, obs}}$ measurements at each frequency.}
    \label{fig:adiab}
\end{figure}

Brightness temperatures as a function of the projected distance from the core can identify the dominant energy loss mechanism leading to the flux decay of jet components \citep{1999ApJ...521..509L, 2005AJ....130.1418J, 2022A&A...660A...1B}. According to the shock-in-jet model described by \citet{1985ApJ...298..114M}, where an adiabatically expanding shock travels downstream in the jet, the main evolutionary stages the component goes through are characterized by Compton, synchrotron and adiabatic losses. In this scenario, brightness temperatures decay as power-laws, T$_{\mathrm{b,jet}}\propto d^{-\epsilon}$, where $d$ is the distance of the jet component from the core, and $\epsilon$ is the power-law index, as the shock moves further away from the core \citep{2012A&A...537A..70S, 2016MNRAS.462.2747K}. In the case of \object{TXS\,1508+572}, jet brightness temperatures can be described with power-law indices of $3.1\pm0.6$ at $15$\,GHz, $-0.3\pm1.11$ at $22$\,GHz, and $-1.7\pm0.3$ at $43$\,GHz (see Fig.~\ref{fig:adiab}). $T_{\mathrm{b}}$ gradients derived for 28 sources at $43$\,GHz by \citet{2022A&A...660A...1B} range from $-3.19$ to $-1.23$, with an average of $-2.07$, so our $43$-GHz power-law index is consistent with these values. This $T_{\mathrm{b}}$ gradient places the jet of \object{TXS\,1508+572} in the Compton loss stage. The inverted and flat gradients at $15$ and $22$\,GHz might be explained via the ongoing activity in the source. We suggest that the $\gamma$-ray activity observed in February 2022 is caused by a shock-shock interaction in the jet region of \object{TXS\,1508+572} and new plasma flowing through the C1 and C2 components. A similar phenomenon was observed in the blazar \object{CTA\,102} by \citet{2011A&A...531A..95F} when a new shock wave traveled through a stationary re-collimation shock.

\section{Summary}
\label{sum}

After exhibiting a bright $\gamma$-ray flare, we started an intensive multi-wavelength follow-up campaign of the early-Universe blazar \object{TXS\,1508+572}. To our knowledge, this is the first attempt at such observations of a flaring high-redshift AGN. While Paper~I discusses the multi-wavelength properties of the source based on the quasi-simultaneous data we collected, here we focused on the VLBI observations included in our monitoring. 

The present study of \object{TXS\,1508+572} extends our knowledge on the evolution of jet geometry and kinematics in high-redshift AGN. Our hybrid images reveal a compact core-jet structure on parsec scales (see Fig.~\ref{fig:cln}). This morphology is affected by the $\gamma$-ray flare, as we recover changes in source structure and brightness on the timescale of months. Jet proper motion values of $0.12-0.27$\,mas/yr are recovered, corresponding to superluminal speeds of 14.3 -- 32.2\,$c$. This result is comparable to the high Lorentz factors of $20$ used to model the SED in Paper I, and is consistent with maximum apparent speeds assuming $\Gamma=20$ in the $(\mu-z)$ relation for high-redshift AGN \citep{2022ApJ...937...19Z}. We trace back the ejection time of the jet component to be between $2016$ and $2019$, during which \object{TXS\,1508+572} was in an elevated state in the $\gamma$-rays. This means that the jet component was not ejected as a result of the high-energy flare in $2022$ February.

Using our multi-frequency data, we measured the distance to the central engine based on the core shift. The distance to the jet apex stays consistent within the measurement errors throughout our observations. On average, the central engine is located $46.1\pm2.3~\mu$as or $0.32\pm0.02$~pc from the $43$-GHz VLBI core. Under the equipartition assumption, which is supported by our brightness temperature measurements in the presence of Doppler boosting, we measure $B^{\mathrm{eq}}_{\mathrm{1pc}}$ of $1.91\pm0.11$~G, $1.57\pm0.17$~G, and $1.79\pm0.16$~G for the three epochs. Low to intermediate redshift AGN also exhibit similar values of $B^{\mathrm{eq}}_{\mathrm{1pc}}$ \citep{2012A&A...545A.113P, 2014Natur.510..126Z}. $B^{\mathrm{non-eq}}_{\mathrm{1pc}}$ values are significantly higher than this, with $324.7\pm156.3$~G, $104.9\pm82.7$~G, and $341.9\pm225.2$~G for the consecutive epochs.

We note that even though we do not observe any brightening in the jet that would be clearly associated with the $\gamma$-ray flare of 2022, there might be a traveling disturbance, such as a density enhancement, that was not emitting much at radio waves but affected the position of the apparent cores at different frequencies during the 2022.67 epoch. This scenario coherently explains our kinematics and core shift measurements. Based on our analysis, we propose that the activity was caused by a shock-shock interaction between the already existing jet component of \object{TXS\,1508+572} and new plasma flowing through this region.

\vspace{1cm}
\begin{acknowledgements}
The authors thank the anonymous referee for useful comments that helped to improve the manuscript. The data were obtained at VLBA within the proposal BG281. The National Radio Astronomy Observatory is a facility of the National Science Foundation operated under cooperative agreement by Associated Universities, Inc. This work made use of the Swinburne University of Technology software correlator, developed as part of the Australian Major National Research Facilities Programme and operated under license.

This research was supported through a PhD grant from the International Max Planck Research School (IMPRS) for Astronomy and Astrophysics at the Universities of Bonn and Cologne. This publication is part of the M2FINDERS project which has received funding from the European Research Council (ERC) under the European Union’s Horizon2020 Research and Innovation Programme (grant agreement No 101018682). FE, JH, MK, and FR acknowledge support from the Deutsche Forschungsgemeinschaft (DFG, grants 447572188, 434448349, 465409577).

\end{acknowledgements}

\begin{appendix}
\onecolumn

\section{Additional figures and tables}

\begin{table}[h]
    \centering
        \caption{Summary of image parameters.}
    \label{tab:cln}
    \begin{tabular}{c c c c c c c c c} 
    \hline\hline
       Epoch & $\nu$\tablefootmark{a} & $S_{\mathrm{tot}}$\tablefootmark{b} & $S_{\mathrm{peak}}$\tablefootmark{c} & $\sigma$\tablefootmark{d}  & $I_{\mathrm{low}}$\tablefootmark{e}  & $b_{\mathrm{maj}}$\tablefootmark{f} & $b_{\mathrm{min}}$\tablefootmark{g} & PA\tablefootmark{h} \\
         &  [GHz]  & [mJy] & [mJy\,beam$^{-1}$] & [mJy\,beam$^{-1}$] & [mJy\,beam$^{-1}$] & [mas] & [mas] & [\degree] \\
    \hline
    2022.24 & $15$ & $240.5$ &  $141.8$ & $0.08$ & $0.28$ & $0.57$ & $0.32$ & $-18$ \\
     & $22$ & $252.1$ & $110.8$ & $0.07$ & $0.22$ & $0.42$ & $0.23$ & $-13$ \\
     & $43$ & $154.0$ & $50.7$ & $0.19$ & $0.53$ & $0.32$ & $0.21$ & $-5$ \\
     & $86$ & $51.2$ & $45.5$ & $0.30$ & $1.37$ & $0.27$ & $0.18$ & $-49$ \\
     2022.67 & $15$ & $258.5$ & $131.4$ & $0.06$ & $0.20$ & $0.53$ & $0.31$ & $-18$ \\
     & $22$ & $220.5$ & $76.5$ & $0.07$ & $0.19$ & $0.39$ & $0.21$ & $-17$ \\
     & $43$ & $179.3$ & $57.6$ & $0.31$ & $0.92$ & $0.22$ & $0.10$ & $-6$ \\
     2023.05 & $15$ & $234.5$ & $112.1$ & $0.12$ & $0.39$ & $0.70$ & $0.32$ & $2$ \\
     & $22$ & $192.6$ & $84.4$ & $0.19$ & $0.59$ & $0.50$ & $0.23$ & $1$ \\
     & $43$ & $188.6$ & $67.6$ & $0.81$ & $2.37$ & $0.23$ & $0.10$ & $-9$ \\
    \hline
    \end{tabular}
        \tablefoot{
    \tablefoottext{a}{Observing frequency.}
    \tablefoottext{b}{Total flux density.}
    \tablefoottext{c}{Peak brightness.}
    \tablefoottext{d}{Rms noise of the image.
    \tablefoottext{e}{Lowest contour level.}
    \tablefoottext{f}{Beam major axis.}
    \tablefoottext{g}{Beam minor axis.}
    \tablefoottext{h}{Beam position angle.}}
    }
\end{table}

\begin{figure}[h!]
\centering
\begin{minipage}{0.49\linewidth}
  \centering
  \begin{subfigure}{\linewidth}
    \includegraphics[width=0.98\linewidth]{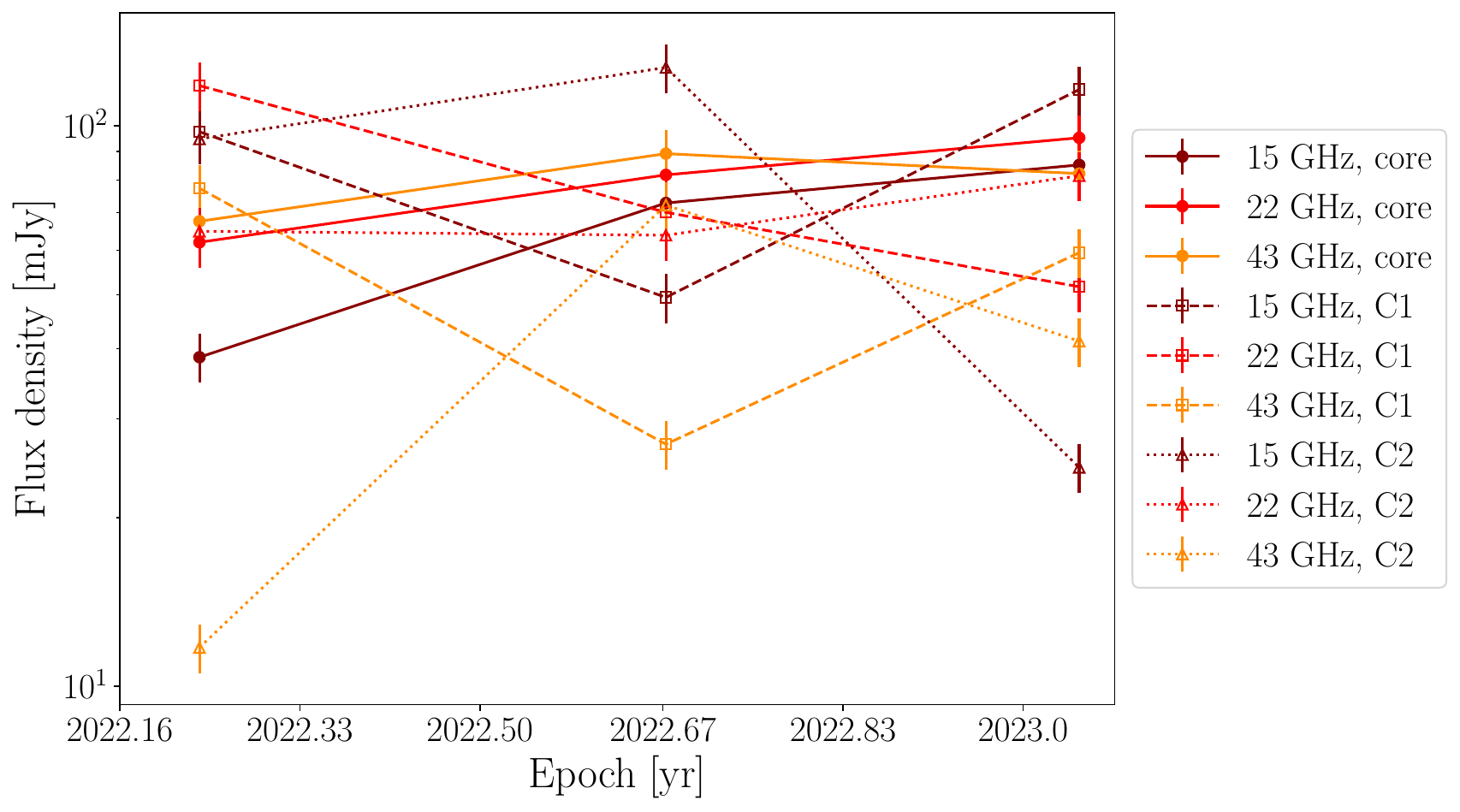}
    \end{subfigure}
    \begin{subfigure}{\linewidth}%
    \includegraphics[width=0.98\linewidth]{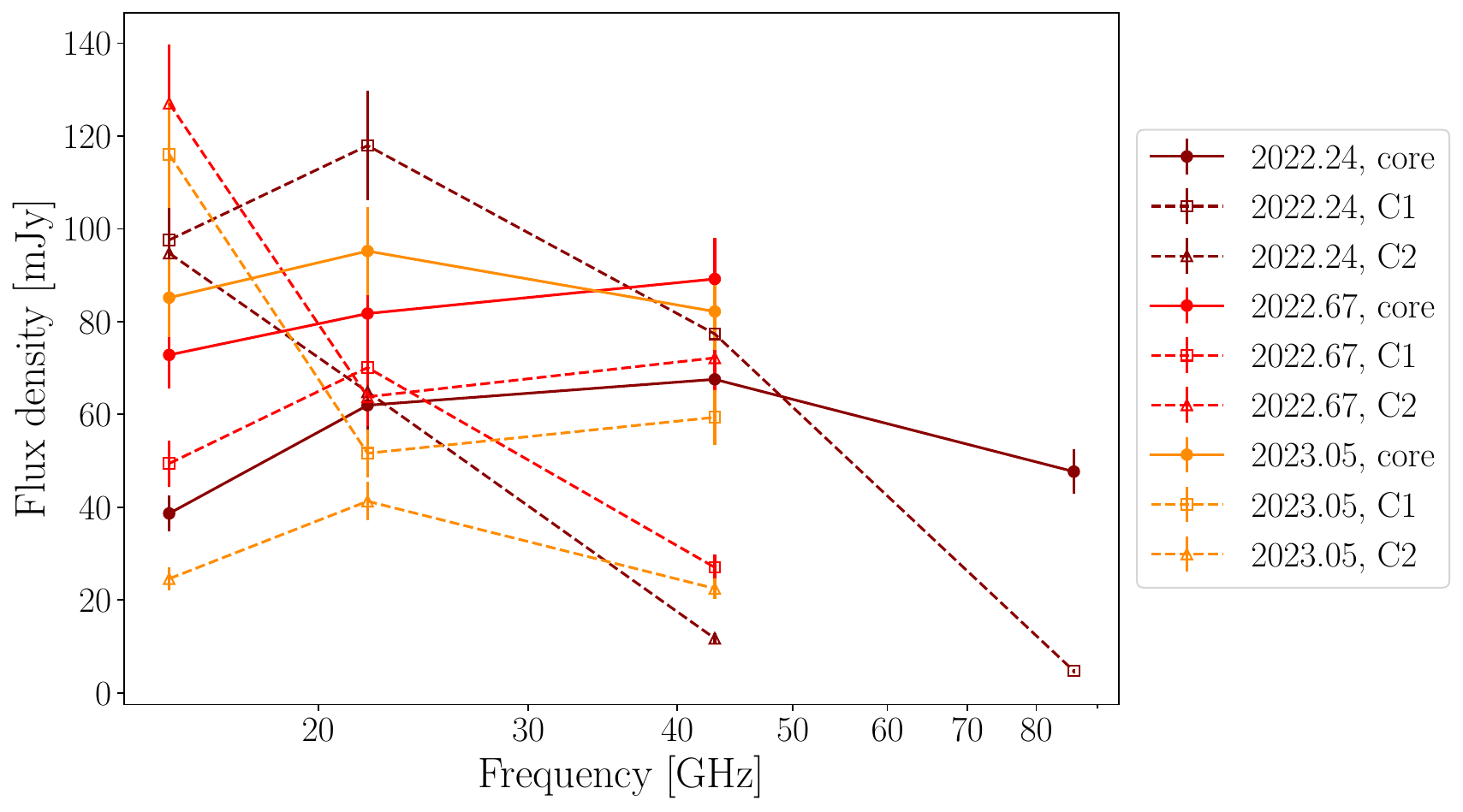}
    \end{subfigure}
    \begin{subfigure}{\linewidth}
    \includegraphics[width=0.98\linewidth]{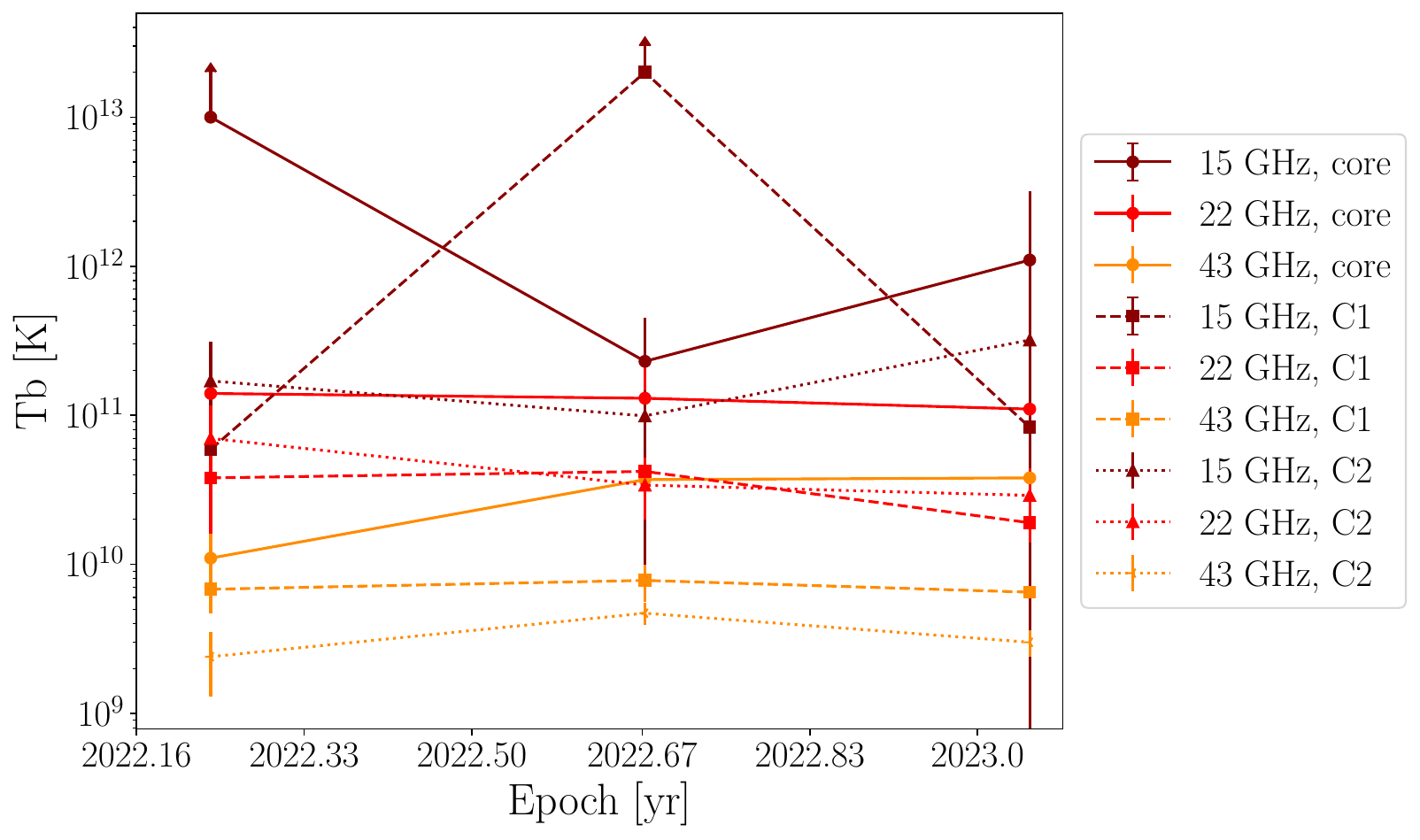}
    \end{subfigure}
  \captionof{figure}{Light curves, spectra, and brightness temperatures of the core and jet components of \object{TXS\,1508+572}.}
  \label{fig:tb}
\end{minipage}
\begin{minipage}{0.5\linewidth}
  \centering
  \includegraphics[width=\linewidth]{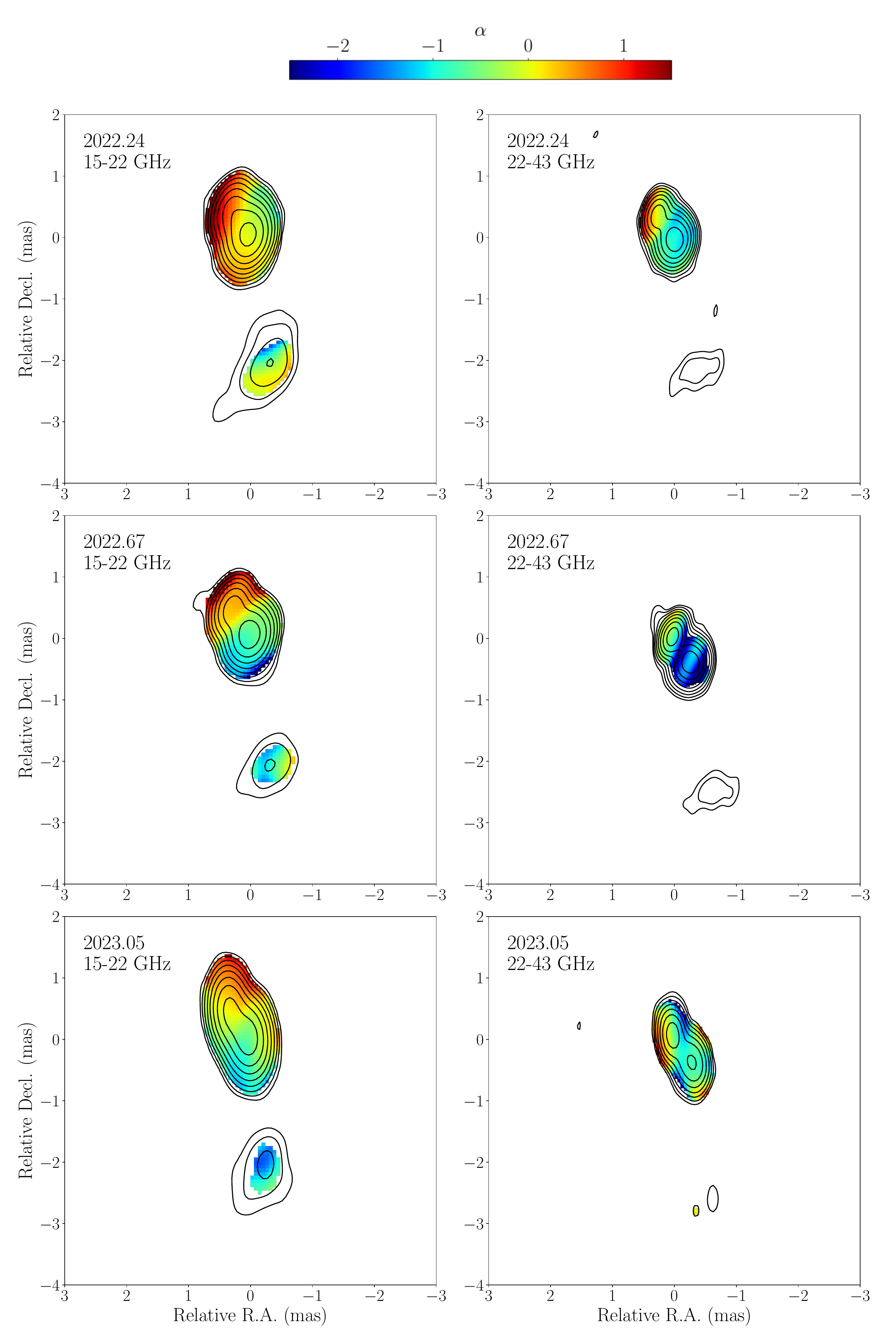}
  \captionof{figure}{Spectral index maps between $15$, $22$ and $43$\,GHz.}
  \label{fig:spix}
\end{minipage}
\end{figure}

\begin{table*}[h]
    \centering
    \caption{Properties of \texttt{modelfit} components.}
    \label{tab:comps}
    \begin{tabular}{c c c c c c c}
    \hline\hline
       $\nu$~[GHz]\tablefootmark{a} & Component & Epoch & $S_{\mathrm{comp}}$~[mJy]\tablefootmark{b} & $b_{\mathrm{maj}}$~[mas]\tablefootmark{c} & $b_{\mathrm{min}}$~[mas]\tablefootmark{d} & $T_{\mathrm{b}}$~[K]\tablefootmark{e} \\
    \hline
        15 & core & 2022.24 & $38.6\pm3.9$ & $<0.01$ & $<0.01$ & $>1.2\times10^{13}$ \\
           &      & 2022.67 & $72.8\pm7.3$ & $0.09\pm0.06$ & $0.09\pm0.06$ & $(2.3\pm2.2)\times10^{11}$ \\
           &      & 2023.05 & $85.1\pm8.5$ & $0.05\pm0.06$ & $0.05\pm0.06$ & $(1.1\pm2.1)\times10^{12}$ \\
           & C1 & 2022.24 & $97.6\pm9.8$ & $0.32\pm0.06$ & $0.14\pm0.06$ & $(5.9\pm3.0)\times10^{10}$ \\
           &      & 2022.67 & $49.4\pm4.9$ & $<0.01$ & $<0.01$ & $>2.0\times10^{13}$ \\
           &      & 2023.05 & $116.1\pm11.6$ & $0.20\pm0.06$ & $0.20\pm0.06$ & $(8.3\pm4.0)\times10^{10}$ \\
           & C2 & 2022.24 & $94.7\pm9.5$ & $0.18\pm0.06$ & $0.09\pm0.06$ & $(1.7\pm1.4)\times10^{11}$ \\
           &      & 2022.67 & $127.1\pm12.7$ & $0.19\pm0.06$ & $0.19\pm0.06$ & $(9.9\pm4.7)\times10^{10}$ \\
           &      & 2023.05 & $24.6\pm2.5$ & $0.05\pm0.06$ & $0.05\pm0.06$ & $(3.2\pm6.7)\times10^{11}$ \\
           & lobe & 2022.24 & $6.6\pm0.7$ & $0.38\pm0.06$ & $0.38\pm0.06$ & $(1.2\pm0.3)\times10^{9}$ \\
           &      & 2022.67 & $6.7\pm0.7$ & $0.48\pm0.06$ & $0.48\pm0.06$ & $(8.1\pm1.7)\times10^{8}$ \\
           &      & 2023.05 & $6.4\pm0.6$ & $0.60\pm0.06$ & $0.60\pm0.06$ & $(4.8\pm1.0)\times10^{8}$ \\
        22 & core & 2022.24 & $62.0\pm6.2$ & $0.07\pm0.05$ & $0.07\pm0.05$ & $(1.4\pm1.3)\times10^{11}$ \\
           &      & 2022.67 & $81.7\pm8.2$ & $0.08\pm0.04$ & $0.08\pm0.04$ & $(1.3\pm1.0)\times10^{11}$ \\
           &      & 2023.05 & $95.2\pm9.5$ & $0.10\pm0.05$ & $0.10\pm0.05$ & $(1.1\pm0.7)\times10^{11}$ \\
           & C1 & 2022.24 & $117.9\pm11.8$ & $0.25\pm0.05$ & $0.15\pm0.05$ & $(3.8\pm1.4)\times10^{10}$ \\
           &      & 2022.67 & $70.1\pm7.0$ & $0.14\pm0.04$ & $0.14\pm0.04$ & $(4.2\pm1.8)\times10^{10}$ \\
           &      & 2023.05 & $51.6\pm5.2$ & $0.18\pm0.05$ & $0.18\pm0.05$ & $(1.9\pm0.7)\times10^{10}$ \\
           & C2 & 2022.24 & $64.8\pm6.5$ & $0.13\pm0.05$ & $0.09\pm0.05$ & $(7.0\pm4.7)\times10^{10}$ \\
           &      & 2022.67 & $63.8\pm6.4$ & $0.15\pm0.04$ & $0.15\pm0.04$ & $(3.4\pm1.4)\times10^{10}$ \\
           &      & 2023.05 & $41.3\pm4.1$ & $0.13\pm0.05$ & $0.13\pm0.05$ & $(2.9\pm1.5)\times10^{10}$ \\
           & lobe & 2022.24 & $6.3\pm0.6$ & $0.55\pm0.05$ & $0.55\pm0.05$ & $(2.4\pm0.4)\times10^{8}$ \\
           &      & 2022.67 & $4.8\pm0.5$ & $0.51\pm0.04$ & $0.51\pm0.04$ & $(2.2\pm0.3)\times10^{8}$ \\
           &      & 2023.05 & $8.0\pm0.8$ & $1.86\pm0.05$ & $1.86\pm0.05$ & $(2.7\pm0.3)\times10^{7}$ \\
        43 & core & 2022.24 & $67.6\pm6.8$ & $0.14\pm0.04$ & $0.14\pm0.04$ & $(1.1\pm0.5)\times10^{10}$ \\
           &      & 2022.67 & $89.2\pm8.9$ & $0.09\pm0.02$ & $0.09\pm0.02$ & $(3.7\pm1.2)\times10^{10}$ \\
           &      & 2023.05 & $82.2\pm8.2$ & $0.09\pm0.02$ & $0.09\pm0.02$ & $(3.8\pm1.3)\times10^{10}$ \\
           & C1 & 2022.24 & $77.3\pm7.7$ & $0.20\pm0.04$ & $0.19\pm0.04$ & $(6.8\pm2.1)\times10^{9}$ \\
           &      & 2022.67 & $27.0\pm2.7$ & $0.11\pm0.02$ & $0.11\pm0.02$ & $(7.8\pm2.2)\times10^{9}$ \\
           &      & 2023.05 & $59.4\pm5.9$ & $0.18\pm0.02$ & $0.18\pm0.02$ & $(6.5\pm1.2)\times10^{9}$ \\
           & C2 & 2022.24 & $11.7\pm1.2$ & $0.14\pm0.04$ & $0.13\pm0.04$ & $(2.4\pm1.1)\times10^{9}$ \\
           &      & 2022.67 & $72.2\pm7.2$ & $0.23\pm0.02$ & $0.23\pm0.02$ & $(4.7\pm0.8)\times10^{9}$ \\
           &      & 2023.05 &  $22.45\pm2.2$ & $0.16\pm0.02$ & $0.16\pm0.02$ & $(3.0\pm0.6)\times10^{9}$ \\
        86 & core & 2022.24 & $47.7\pm4.8$ & $0.04\pm0.04$ & $0.04\pm0.04$ & ($3.0\pm5.0)\times10^{13}$ \\
           & C1  & 2022.24 & $4.7\pm0.5$ & $0.08\pm0.04$ & $0.08\pm0.04$ & ($6.0\pm6.0)\times10^{11}$ \\
    \hline
    \end{tabular}
    \tablefoot{
    \tablefoottext{a}{Observing frequency.}
    \tablefoottext{b}{Flux density.}
    \tablefoottext{c}{Component major axis.}
    \tablefoottext{d}{Component minor axis.}
    \tablefoottext{e}{Brightness temperature corrected for redshift.}
    }
\end{table*}

\twocolumn
\section{Multi-frequency kinematic fit}
\label{mf_kin}

\begin{figure}[h!]
    \centering
    \includegraphics[width=\linewidth]{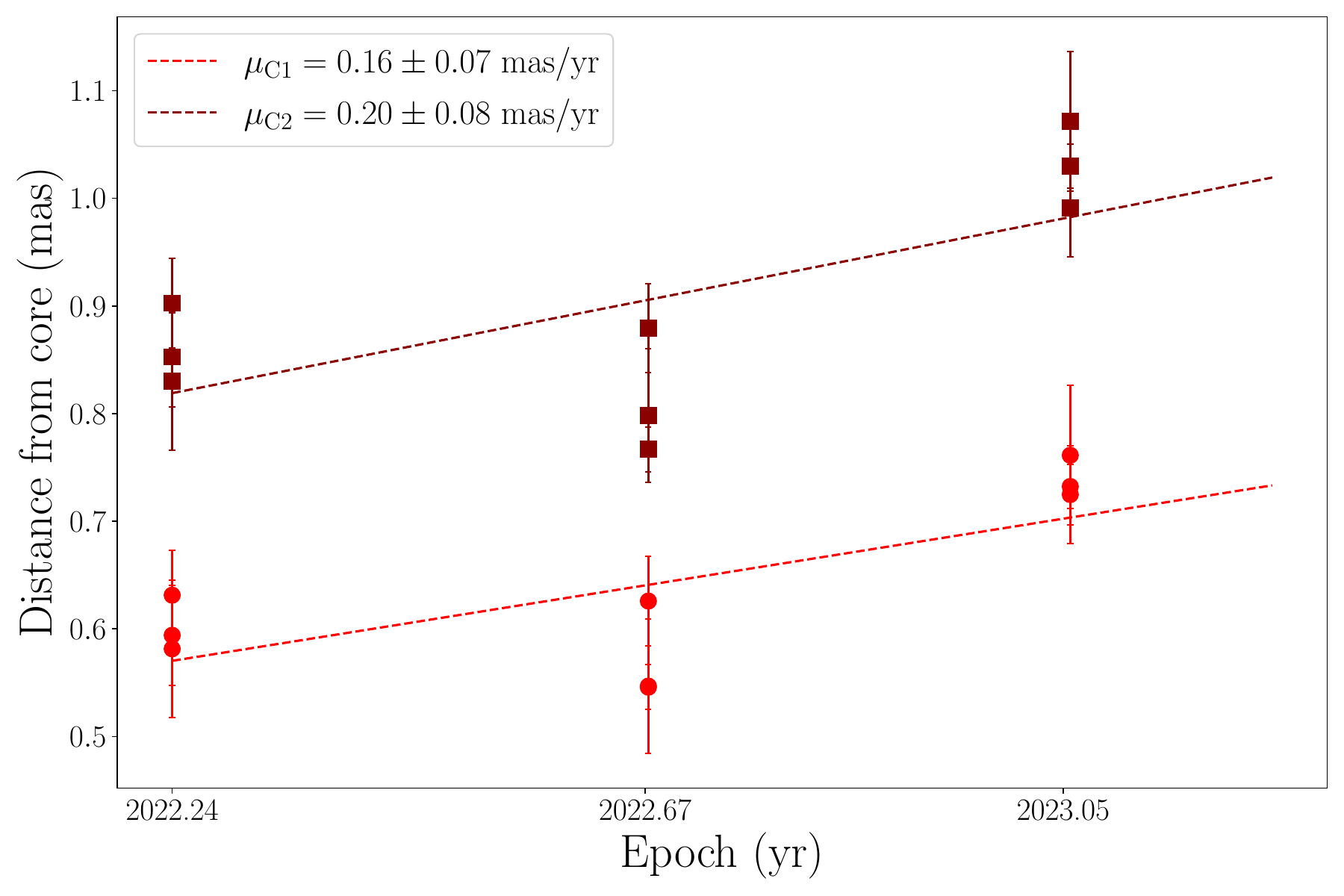}
    \caption{Kinematics of the jet components C1 (red circles) and C2 (dark red squares), with their positions aligned based on the core shift measurement.}
    \label{fig:mfkin}
\end{figure}

Here we present a multi-frequency kinematic analysis of the two jet components whose positions were aligned based on the core shift measurement (see Sect.~\ref{spec_cs}). The component distances and the best fit functions are shown in Fig.~\ref{fig:mfkin}. Our data is well described with a linear fit which reveals apparent component speed of $\mu_{\mathrm{C1}}=0.16\pm0.07$~mas/yr and $\mu_{\mathrm{C2}}=0.20\pm0.08$~mas/yr.

However, the amount of data available for kinematic estimates is not sufficient to justify higher orders of trajectory fit, and our overall conclusions are not strongly dependent on the linearity of the apparent trajectory. In addition, according to the methodology established for AGN monitoring data \citet{2009ApJ...706.1253H}, component acceleration is only analyzed if a given jet feature is robustly detected in at least ten observing epochs. Linear fits are more suitable to determine the component speed and ejection time. Nevertheless, our data presented here would be a useful set of future study of kinematics which might favor a higher order of trajectory fit.

\end{appendix}

\end{document}